\documentclass[11pt]{article}

\usepackage[final]{acl}

\usepackage{times}
\usepackage{latexsym}
\usepackage{booktabs}
\usepackage{multirow}
\usepackage{algpseudocode}
\usepackage{colortbl}
\usepackage{array}
\usepackage{xcolor}
\usepackage{amsmath}
\usepackage{bm}
\usepackage{amsfonts}
\usepackage{amssymb}
\definecolor{lightgray}{gray}{0.88}
\usepackage{pifont}
\usepackage{threeparttable}
\usepackage{ulem}      
\definecolor{verylightgray}{gray}{0.92}
\newcommand{\secondbest}[1]{%
	\cellcolor{verylightgray}\uline{#1}%
}
\usepackage{caption}

\usepackage[T1]{fontenc}

\usepackage[utf8]{inputenc}

\usepackage{microtype}

\usepackage{inconsolata}

\usepackage{graphicx}

\title{RTCFake: Speech Deepfake Detection in Real-Time Communication}

\author{Jun Xue\textsuperscript{1,2}, Zhuolin Yi\textsuperscript{1,2}, Yihuan Huang\textsuperscript{1,2}, Yanzhen Ren\textsuperscript{1,2}\thanks{Corresponding author}, Yujie Chen\textsuperscript{4}, Cunhang Fan\textsuperscript{3} \\
	\textbf{Zicheng Su\textsuperscript{1,2}, Yongcheng Zhang\textsuperscript{1,2}, Bo Cai\textsuperscript{1,2}}  \\
  \textsuperscript{1}Key Laboratory of Aerospace Information Security and Trusted Computing, Ministry of Education  \\
  \textsuperscript{2}School of Cyber Science and Engineering, Wuhan University, Wuhan, China \\
  \textsuperscript{3}School of Computer Science and Technology, Anhui University, Hefei, China \\
   \textsuperscript{4}Beihang University, Beijing, China \\
  \texttt{\{junxue,yizhuolin,renyz\}@whu.edu.cn}, cunhang.fan@ahu.edu.cn \\
}

\begin{document}
\maketitle
\begin{abstract}
With the rapid advancement of speech generation technologies, the threat posed by speech deepfakes in real-time communication (RTC) scenarios has intensified. However, existing detection studies mainly focus on offline simulations and struggle to cope with the complex distortions introduced during RTC transmission, including unknown speech enhancement processes (e.g., noise suppression) and codec compression. To address this challenge, we present the first large-scale speech deepfake dataset tailored for RTC scenarios, termed \textit{RTCFake}, totaling approximately 600 hours. The dataset is constructed by transmitting speech through multiple mainstream social media and conferencing platforms (e.g., Zoom), enabling precise pairing between offline and online speech. In addition, we propose a phoneme-guided consistency learning (PCL) strategy that enforces models to learn platform-invariant semantic structural representations. In this paper, the RTCFake dataset is divided into training, development, and evaluation sets.  The evaluation set further includes both unseen RTC platforms and unseen complex noise conditions, thereby providing a more realistic and challenging evaluation benchmark for speech deepfake detection. Furthermore, the proposed PCL strategy achieves significant improvements in both cross-platform generalization and noise robustness, offering an effective and generalizable modeling paradigm. The \textit{RTCFake} dataset is provided in the {\url{https://huggingface.co/datasets/JunXueTech/RTCFake}}.

\end{abstract}

\section{Introduction}
The rapid evolution of text-to-speech (TTS)  \cite{zhou2025voxcpm} and voice conversion (VC) \cite{du2024cosyvoice} technologies has drastically lowered the barrier for high-fidelity speech synthesis. With the prevalence of online conferencing and remote collaboration, voice has become a cornerstone of identity verification in online interactions. A prime example is a 2025 incident where a corporate executive was targeted in a \$499,000 fraud during a Zoom meeting featuring an AI-impersonated CEO \cite{cna_deepfake_scam_2025}. Such deepfake threats have posed a severe risk to the security of Real-Time Communication (RTC).

\begin{figure}[!t]  
		\vspace{-16pt}
	\centering
	\includegraphics[scale=0.65]{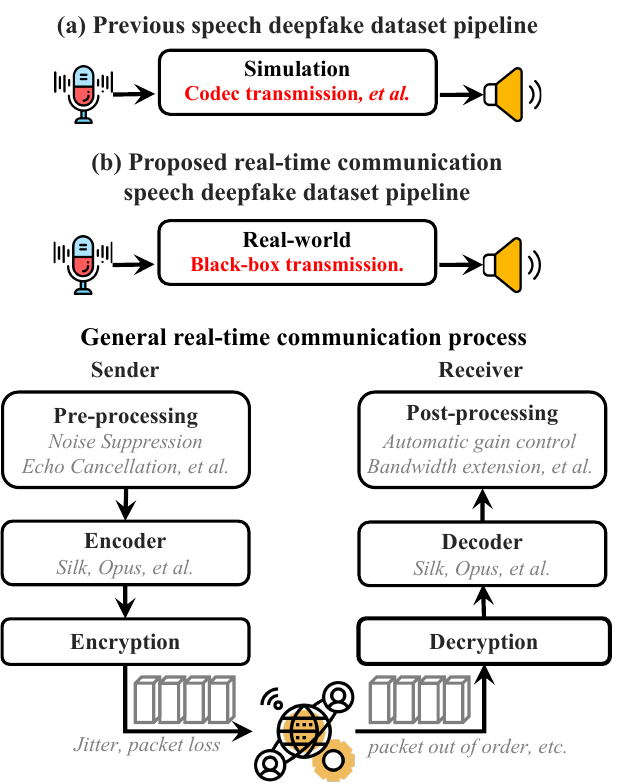}
	\caption{Illustration of different speech deepfake dataset construction pipelines.
		(a) Previous datasets mainly rely on simulated distortions, such as codec-based transmission, to model communication effects.
		(b) The proposed \textit{RTCFake} dataset is constructed under real-world black-box transmission conditions, where communication over mainstream social media platforms typically involves key distortion sources, including noise suppression, echo cancellation, codec encoding and decoding, and packet loss during network transmission.}
	
	\label{fig:rtc}
		\vspace{-16pt}
\end{figure}

\begin{figure*}[!t]  
	\centering
	\includegraphics[scale=0.45]{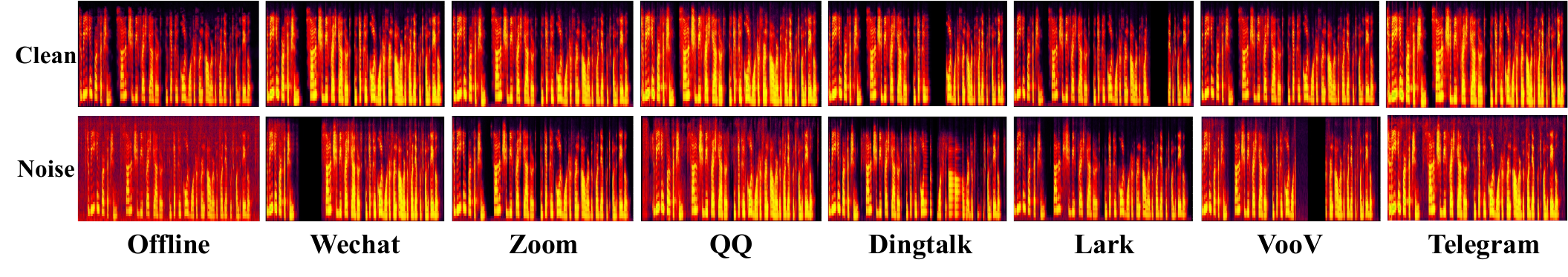}
	\caption{Amplitude spectrograms of the same utterance under offline and multiple online transmission conditions. The top row shows the clean speech, while the bottom row shows the corresponding noisy speech.}
	
	\label{fig:spectro}
\end{figure*}

Recently, the research community has introduced a series of Synthetic Speech Detection (SSD) challenges to safeguard the authenticity of voice interactions, gaining significant attention. For instance, ASVspoof 2021 \cite{yamagishi2021asvspoof} and ASVspoof 5 \cite{wang2024asvspoof}  highlighted voice distortions caused by codecs and lossy compression (e.g., MP3), while ADD 2022 \cite{yi2022add} incorporated various real-world noises. However, factors in real-world voice interactions are highly coupled and dynamically evolving. Existing challenges often focus on isolated factors, making it difficult to faithfully evaluate model performance in authentic and complex environments.

Mainstream RTC platforms typically exhibit \emph{black-box} characteristics, in which speech undergoes a series of highly integrated processing modules (as illustrated in Fig.~\ref{fig:rtc}). On the sender side, speech is first processed by front-end modules such as noise suppression and echo cancellation, followed by encoding and encrypted transmission. During network transmission, the signal is inevitably affected by jitter and packet loss. Finally, on the receiver side, the speech is decrypted, decoded, and further processed through post-processing modules such as gain control and bandwidth extension. These modules are tightly coupled in both the temporal and spectral domains, leading to systematic and nonlinear alterations in the energy distribution and acoustic structure of the transmitted speech (as shown in Fig.~\ref{fig:spectro}).

Such realistic and complex RTC environments pose multiple critical challenges for SDD.
\textbf{(1) Black-box processing:} As the transmission strategies of RTC platforms are unknown, detection models are required to capture forgery cues under black-box conditions.
\textbf{(2) Noise robustness challenges:} Built-in enhancement algorithms in RTC systems (e.g., noise suppression and echo cancellation), while improving perceptual speech quality, often suppress or distort fine-grained forgery artifacts in synthetic speech.
\textbf{(3) Cross-platform generalization challenges:} Due to variations in front-end processing logic and transmission protocols across platforms, the distortion distributions imposed on speech signals differ significantly.

\begin{table*}[t]
	\centering
	\begin{threeparttable}
	\caption{Overview of the RTCFake dataset and comparison with existing speech deepfake datasets.}
	\setlength{\tabcolsep}{2pt}
	\label{tab:dataset_compare}
	\begin{tabular}{l c c p{6cm}}
		\specialrule{1.2pt}{0pt}{0pt}
		\textbf{Dataset} & \textbf{Year} & \textbf{Duration$^{*}$}
		 & \textbf{Description} \\
		\midrule
		ASVspoof2019 \cite{todisco2019asvspoof} & 2019 & 116 & Clean scenario\\
		ASVspoof2021 \cite{liu2023asvspoof} & 2021 & -- &  Codec transmission data\\
		ADD2022 \cite{yi2022add} & 2022 & -- & Noise scenario \\
		ADD2023 \cite{yi2023add} & 2023 & -- & Generation–detection adversarial\\
		ASVspoof5 \cite{wang2024asvspoof} & 2024 & 600 & Crowdsourced data \\
		CD-ADD \cite{li2024cross} & 2024 & 384 & Cross-domain deepfake  \\
		MLAAD \cite{muller2024mlaad} & 2024 & -- & Multi-language data \\
		DFADD \cite{du2024dfadd} & 2024 & 200 & Based on diffusion and flow-matching  \\
		CVoiceFake \cite{li2024safeear} & 2024 & -- & Content privacy-preserving\\
		FSW \cite{xie2025fake} & 2025 & 254 & Fake speech in-the-wild\\
		SpoofCeleb \cite{jung2025spoofceleb} & 2025 & 1982 & In the wild \\
		CodecFake \cite{xie2025codecfake} & 2025 & -- & Based on audio language model \\
		SpeechFake \cite{wen2025speechfake} & 2025 & 3000 & Multi-language data\\
		\midrule
		\rowcolor{gray!15}
		\textbf{RTCFake (Ours)} & 2025 & 600 & \textbf{Real-time Communication}\\
		\specialrule{1.2pt}{0pt}{0pt}
	\end{tabular}
\begin{tablenotes}
	\footnotesize \item $^{*}$Duration is measured in hours.
\end{tablenotes}
\end{threeparttable}
\end{table*}

To address the above challenges, we introduce the RTCFake dataset, a large-scale speech deepfake dataset specifically constructed for RTC scenarios. The dataset is built by first generating offline data and then transmitting it through mainstream RTC platforms, totalling approximately 600 hours. Based on \textit{RTCFake}, we analyze offline and online speech representations, and observe that phoneme-level representations exhibit substantially higher stability than frame-level representations. Therefore, we propose a phoneme-guided consistency learning strategy, which encourages the model to focus on discriminative features that remain invariant across offline-to-online transformations during training, thereby effectively mitigating the adverse impact of RTC-induced distortions.

The main contributions of this work are summarized as follows:
\begin{itemize}
		\vspace{-5px}
	\item \textbf{\textit{RTCFake} Dataset:} We construct the first large-scale speech deepfake dataset tailored for RTC scenarios, comprising 600 hours of data transmitted through mainstream communication platforms, which provides a foundation for studying speech deepfake detection under real-world communication conditions.
	\vspace{-5px}
	\item \textbf{Phoneme-Guided Consistency Learning:} Leveraging the high stability of phoneme-level representations under communication transmission, we propose a phoneme-guided consistency learning strategy that constrains the model to focus on discriminative features that remain invariant across offline--online scenarios during training.
	\vspace{-5px}
	\item \textbf{Robustness and Generalization Evaluation:} Extensive experiments conducted under multiple noise conditions and cross-platform evaluation settings demonstrate that the proposed dataset and method significantly improve the robustness and generalization performance of detection models in realistic RTC scenarios.
	\vspace{-4px}
\end{itemize}

\section{Related Work}

\paragraph{Speech Deepfake Detection Datasets.} In recent years, several key benchmark datasets have emerged in the field of SDD. As summarized in Table~\ref{tab:dataset_compare}, the ASVspoof series \cite{todisco2019asvspoof, liu2023asvspoof, wang2024asvspoof} established standardized evaluation protocols, with ASVspoof 2021 and ASVspoof 5 incorporating channel codec effects. Subsequently, datasets such as ADD 2023 \cite{yi2023add}, DFADD \cite{du2024dfadd}, CodecFake \cite{xie2025codecfake}, and SpeechFake \cite{wen2025speechfake} introduced high-fidelity synthetic speech based on diffusion models, flow matching, and modern codecs. Furthermore, SpoofCeleb \cite{jung2025spoofceleb} and FakeSpeechWild \cite{xie2025fake} explored the complexities of unconstrained real-world conditions by collecting samples from open platforms.

However, these datasets have not yet accounted for the highly coupled and dynamically evolving black-box processing pipelines inherent in RTC scenarios. To bridge this gap, we construct the RTCFake dataset. By transmitting utterances through mainstream communication platforms to obtain paired "offline-online" speech, RTCFake provides solid data support in speech interaction scenarios.

\paragraph{Speech Deepfake Detection Methods.} Existing speech deepfake detection methods primarily focus on mining key discriminative cues from speech signals. Current research typically relies on handcrafted acoustic features \cite{fan2024spatial,xue2022audio}, end-to-end deep models \cite{jung2022aasist,xue2023learning,xue2024dynamic}, and self-supervised speech representation learning \cite{zhang2024audio}. Meanwhile, robust modeling against noise interference \cite{fan2024dual} and codec-related distortions \cite{wu2024codecfake} has also received significant attention.

However, most existing methods rely on frame-level acoustic features. In RTC scenarios, these subtle frame-level cues are easily erased by complex nonlinear processing modules. Due to the lack of effective modeling for transmission-invariant features across platforms, existing strategies face severe performance challenges in real-world communication environments.

\begin{figure*}[!t]  
	\centering
	\includegraphics[scale=0.54]{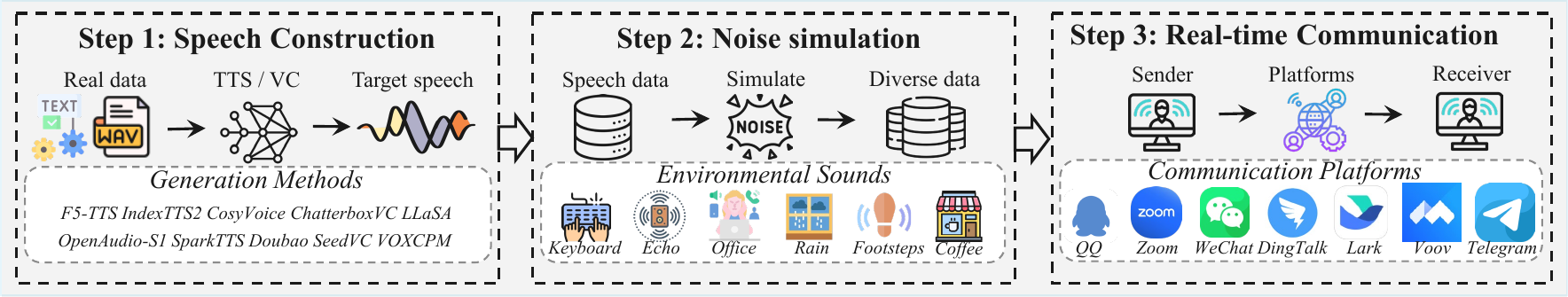}
	\caption{Offline data construction and real-time communication  transmission pipeline}
	\label{fig:main}
\end{figure*}

\section{Dataset Collection and Statistics}
This section describes the overall construction and transmission pipeline of the \textit{RTCFake} dataset, as illustrated in Fig.~\ref{fig:main}. The pipeline consists of three successive stages: offline speech construction, noise simulation, and RTC transmission. By integrating diverse speech generation methods, representative environmental disturbances, and end-to-end transmission through mainstream communication platforms, we construct paired offline--online speech data. This dataset provides a solid foundation for conducting SDD research in realistic communication scenarios.

\subsection{Speech Construction}
The offline dataset is constructed from both real and synthetic speech. Real speech is collected from open-source corpora, with English data sourced from LibriHeavy \cite{kang2024libriheavy} and Chinese data from Chinese-Lips \cite{zhao2025chinese}, ensuring diversity in natural speech across languages and speaking styles. Synthetic speech is generated using a variety of mainstream TTS and VC systems, as summarized in Table~\ref{tab:gen_platform}. In total, we employ seven TTS systems (G01–G07) and three VC systems (G08–G10), covering a wide range of representative modeling paradigms in contemporary speech generation research.

As shown in Table~\ref{tab:offline_dataset_overview}, the offline dataset is divided into training, development, and evaluation subsets. To provide a basis for subsequent analysis of processing mechanisms such as noise suppression and echo cancellation during real-time transmission, multiple scenario-specific noises are introduced into the evaluation set, including office and coffee\footnote{https://media.xiph.org/rnnoise/data/}, echo\footnote{https://github.com/CLAD23/CLAD/tree/main}, and three types of environmental noise \cite{piczak2015esc}.

\subsection{Real-time Communication}
To simulate realistic RTC scenarios, we employ two independent PC devices serving as the sender and the receiver, respectively. On the sender side, offline speech samples are played through a virtual audio device and fed into mainstream RTC platforms for voice transmission; on the receiver side, the incoming audio streams are captured and recorded in real time using OBS\footnote{https://obsproject.com/}. The transmission process covers multiple widely used communication platforms, including Wechat, Zoom, QQ, DingTalk, Lark, VooV, and Telegram. To improve transmission efficiency, a subset of speech samples is concatenated prior to transmission; after transmission, the received audio is segmented based on timestamp information to recover individual utterances.

In addition, automatic speech recognition (ASR) \cite{radford2023robust} is used to verify the consistency between the transcribed content of the transmitted speech and the original labels, and samples with mismatched content are discarded. Through this procedure, we obtain online speech samples that closely reflect real user communication behavior while maintaining reliable annotations.

\subsection{Dataset Statistics}
The RTCFake dataset consists of two subsets, offline and online, with a total duration of approximately 600 hours and coverage of 307 speakers. Table \ref{tab:dataset_compare} presents a comparison between RTCFake and existing speech deepfake datasets. For a more detailed introduction to the dataset, please refer to Appendix Section \ref{sec:appendix}.

\section{Methodology}

\subsection{Analysis and Motivation}

The distortions introduced within RTC systems exhibit distinct hierarchical characteristics. Processing modules such as codec compression and noise suppression severely perturb local temporal structures and instantaneous acoustic details, leading to a pronounced distribution shift in frame-level representations between offline and online speech. Nevertheless, these frame-level features often contain critical fine-grained artifacts essential for SDD. 

\begin{figure}[h]  
	\centering
	\includegraphics[scale=0.38]{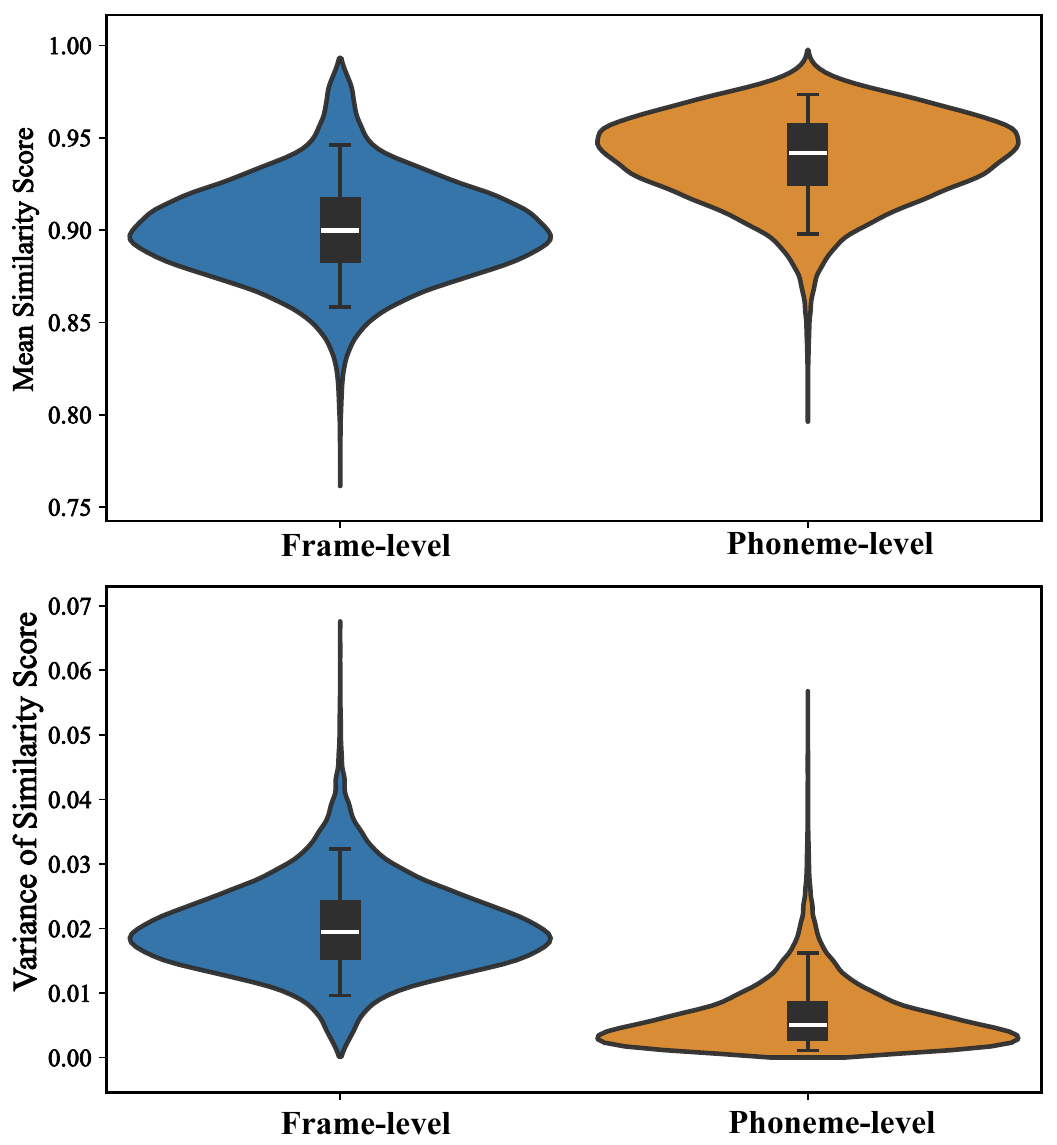}
	\caption{Comparison of offline--online representation similarity at frame-level \cite{babu2022xls} and phoneme-level \cite{xu2022simple} on the RTCFake training set. Phoneme-level representations show higher mean similarity and lower variance, indicating better stability.}
	\label{fig:stable}
\end{figure}

Fig.\ref{fig:stable} presents a comparison of similarity metrics for frame-level and phoneme-level representations before and after transmission, based on paired offline–online utterances from the RTCFake training set. Statistical results demonstrate that phoneme representations possess significantly higher stability than frame representations, characterized by a larger mean similarity and a substantially smaller variance. This phenomenon suggests that RTC systems prioritize the preservation of semantic intelligibility over acoustic fidelity. Consequently, as the fundamental units of linguistic content, phonemes maintain superior consistency during cross-platform transmission.

Motivated by these findings, we propose a phoneme-guided consistency learning approach during training. Within this framework, phoneme-level representations act as robust anchors for cross-domain modeling, encouraging the neural network to learn RTC platform-invariant feature representations, thereby improving the model’s generalization performance.

\begin{table*}[t]
		\begin{threeparttable}
	\centering
	\caption{EER (\%) results under different training datasets and evaluation conditions. The best and second-best results in each column are highlighted in bold and underlined, respectively. Off and On denote models trained on offline and online data, respectively, while Mix indicates training on their combination.
		 Details of P0X conditions are provided in Table~\ref{tab:gen_plat_ID}.}
	\label{tab:eer_eval}
	\begin{tabular}{c|c|c|c|c|c|c|c|c|c|c}
		\specialrule{1.2pt}{1pt}{1pt}
		\multirow{3}{*}{\textbf{Train Data}} 
		& \multicolumn{10}{c}{\textbf{Eval Data (EER $\downarrow$)}} \\
		\cline{2-11}
		& \multirow{2}{*}{Offline}
		& \multicolumn{8}{c|}{Online}
		& \multirow{2}{*}{All} \\
		\cline{3-10}
		& 
		& P01 & P02 & P03 & P04 & P05 & P06 & P07 & avg
		& \\
		\hline 
		
		ASVspoof2019 & 51.15 & 54.68 & 29.70 & 49.71 & 53.87 & 49.45 & 48.23 & 43.67 & 49.40 & 50.28 \\
		DFADD        & 47.86 & 50.45 & 39.49 & 46.70 & 48.38 & 47.73 & 48.26 & 46.42 & 47.56 & 47.71 \\
		ASVspoof5    & 42.49 & 48.76 & 29.99 & 44.09 & 46.86 & 45.72 & 44.88 & 44.66 & 44.92 & 43.71 \\
		FSW          & 43.81 & 38.38 & 58.19 & 43.02 & 40.94 & 43.52 & 44.52 & 43.50 & 43.55 & 43.68 \\
		CodecFake    & 42.25 & 52.90 & 23.74 & 39.96 & 47.55 & 40.11 & 43.59 & 40.33 & 41.55 & 41.90 \\
		speechfake-BD & 35.36 & 47.48 & 24.16 & 38.17 & 44.00 & 37.64 & 38.02 & 40.38 & 40.01 & 37.69 \\
		SpoofCeleb   & 29.56 & 40.05 & 30.70 & 35.16 & 38.54 & 36.41 & 32.48 & 40.33 & 38.55 & 34.06 \\
		CD-ADD       & 33.86 & 42.13 & 26.66 & 32.11 & 40.32 & 35.10 & 34.07 & 31.57 & 34.01 & 33.95 \\
		
		\hline
		
		Off &  \secondbest{5.42} 
		& 6.79 & 20.40 & 13.10 & 12.56 & 16.72 & 16.07 & 19.05 & 13.79 & 9.60 \\
		
		On  & 9.57 
		&5.05
		&\secondbest{7.30 } 
		&  \secondbest{8.05} 
		& 8.79 
		&  \secondbest{10.53} 
		& 11.77 
		&  \secondbest{11.80} 
		&  \secondbest{8.35} 
		& 8.96 \\
		
		Mix     & 6.09 
		&   \secondbest{4.93} 
		& 8.85 
		& 8.10 
		&   \secondbest{8.53} 
		& 10.97 
		&  \secondbest{11.65} 
		& 12.18 
		& 8.57 
		&   \secondbest{7.33} \\
		
	\cellcolor{lightgray}\textbf{PCL$^{*}$}    &\cellcolor{lightgray}\textbf{4.84}
&\cellcolor{lightgray}\textbf{3.79}
&\cellcolor{lightgray}\textbf{6.24}
&\cellcolor{lightgray}\textbf{7.03}
&\cellcolor{lightgray}\textbf{6.76} 
&\cellcolor{lightgray}\textbf{8.51}
&\cellcolor{lightgray}\textbf{10.17} 
&\cellcolor{lightgray}\textbf{8.75}
&\cellcolor{lightgray}\textbf{6.77}
&\cellcolor{lightgray}\textbf{5.81} \\

		\specialrule{1.2pt}{0pt}{0pt}
	\end{tabular}
	\begin{tablenotes}
		\footnotesize \item $^{*}$PCL denotes a phoneme-guided consistency learning strategy trained on paired offline--online data.
		
	\end{tablenotes}
	\end{threeparttable}
\end{table*}

\subsection{Phoneme-guided Consistency Learning}

Based on the above observations, we propose a phoneme-guided consistency learning strategy, which aims to enforce invariance between offline and online representations at the level of semantic structural units.

First, the offline and online speech signals are processed by a shared feature extractor to obtain frame-level acoustic representations. Subsequently, a phoneme recognition model\footnote{https://huggingface.co/facebook/wav2vec2-xlsr-53-espeak-cv-ft} is employed to predict frame-level phoneme boundaries, based on which consecutive frames are aligned into linguistically meaningful phoneme segments. 

Specifically, let $\bm{H} = [{h}_1, {h}_2, \dots, {h}_T]$ denote the sequence of frame-level features. The phoneme-level representation $\bm{p}_k$ is computed via temporal average pooling:
\begin{equation}\bm{p}_k = \frac{1}{|{f}k|} \sum{t \in {f}_k} {h}_t,\end{equation}
where $|{f}_k|$ denotes the number of frames within the $k$-th phoneme boundary. This aggregation is applied to both offline and online feature sequences, resulting in paired phoneme-level representations, denoted as $\bm{p}^{(a)}$ and $\bm{p}^{(b)}$, respectively.

To enhance representation consistency under different transmission conditions, we introduce a bidirectional consistency constraint between paired phoneme-level features $\bm{p}^{(a)}$ and $\bm{p}^{(b)}$. Specifically, consistency is enforced by minimizing the mean squared error (MSE) between the two representations. Accordingly, the phoneme-level consistency learning loss is defined as:
\begin{equation}
	\footnotesize
	\begin{aligned}
		\mathcal{L}_{pcl}( \bm{p}^{(a)}, \bm{p}^{(b)}) = &  D_{\text{MSE}} (p^{(a)} \| p^{(b)}),
	\end{aligned}
\end{equation}

During training, we jointly optimize the classification loss and the phoneme-level consistency learning term to ensure both discriminative capability and representation consistency across different transmission conditions. Specifically, the offline and online branches produce prediction logits, denoted as $\bm{z}^{(a)}$ and $\bm{z}^{(b)}$, respectively, each of which is supervised by the ground-truth label $\bm{y}$ via the cross-entropy loss $\mathcal{L}_{\mathrm{ce}}$. 

By combining the above loss terms, the overall training objective is defined as
\begin{equation}
	\footnotesize
	\begin{aligned}
		\mathcal{L} =\;
		& \frac{1}{2} \big(
		\mathcal{L}_{\mathrm{ce}}(\bm{z}^{(a)}, \bm{y})
		+
		\mathcal{L}_{\mathrm{ce}}(\bm{z}^{(b)}, \bm{y})
		\big) \\
		& + \lambda \, \mathcal{L}_{\mathrm{pcl}}(\bm{p}^{(a)}, \bm{p}^{(b)}),
	\end{aligned}
\end{equation}

where $\lambda$ is a weighting coefficient that balances the classification objective and the phoneme-level consistency learning term.


\section{Experiments and Analysis}

\subsection{Experimental Setup}

To evaluate deepfake speech detection performance, we adopt a state-of-the-art model, XLSR+AASIST \cite{tak2022automatic}. Specifically, XLSR \cite{babu2022xls} serves as the front-end feature extractor, while AASIST employs a heterogeneous stacked graph attention network as the back-end classifier for deepfake detection. During training, we apply RawBoost \cite{tak2022rawboost} to improve robustness. For phoneme representation, we employ the pre-trained Wav2Vec2-Large-XLSR-53 model \cite{xu2022simple} to identify phoneme boundaries. The model is optimized using Adam with a learning rate of $1\times10^{-6}$ and a weight decay of $1\times10^{-4}$. Training is conducted for up to 100 epochs with an early stopping strategy, where training is terminated if no performance improvement is observed for 10 consecutive epochs. During evaluation, we use the Equal Error Rate (EER) as the performance metric. All experiments are conducted on an NVIDIA RTX 4090 GPU.

\subsection{Overall Performance}

Table~\ref{tab:eer_eval} reports EER results under different training datasets and evaluation conditions. The first eight rows correspond to models trained on existing open-source datasets, while the last four rows present models trained on the offline data (Off), online data (On), and mixed data (Mix) constructed in this work, as well as the results obtained after applying the proposed phoneme-guided consistency learning (PCL) strategy.

First, models trained on open-source datasets exhibit very limited generalization capability. They produce extremely high EERs on both the offline test set and multiple online platforms. This indicates that existing open-source datasets fail to capture the complex distributions encountered in real-world applications.

Second, a severe domain mismatch exists between offline and online scenarios. Models trained exclusively on offline data (Off) achieve reasonable performance on the offline test set (5.42\%) but suffer dramatic degradation under online conditions, where the average EER increases to 13.79\%. This clearly demonstrates that black-box transmission in RTC scenarios poses a substantial challenge to spoofing detection. Models trained exclusively on online data (On) improve robustness in online environments; however, their performance on offline data deteriorates markedly, with the EER rising to 9.57\%. This suggests that purely online training introduces domain-specific biases that weaken the model’s ability to detect spoofed speech in offline conditions.

Mixed training (Mix) partially alleviates the domain mismatch by jointly modeling offline and online data, resulting in more balanced performance and a reduced average EER of 7.33\%.
Finally, the proposed PCL method achieves the best overall performance. Compared with the mixed training strategy, PCL consistently reduces EER across all evaluation conditions, achieving the lowest average EER of 5.81\%.

In summary:
(1) Existing open-source datasets are insufficient for evaluating speech deepfake detection under realistic RTC conditions.
(2) RTC transmission induces a severe domain mismatch between offline and online scenarios.
(3) Phoneme-guided consistency learning (PCL) effectively exploits cross-scenario stable characteristics and substantially mitigates domain mismatch under RTC conditions.

\subsection{Cross-Platform Generalization}

\begin{figure}[!t]  
	\centering
	\includegraphics[scale=0.35]{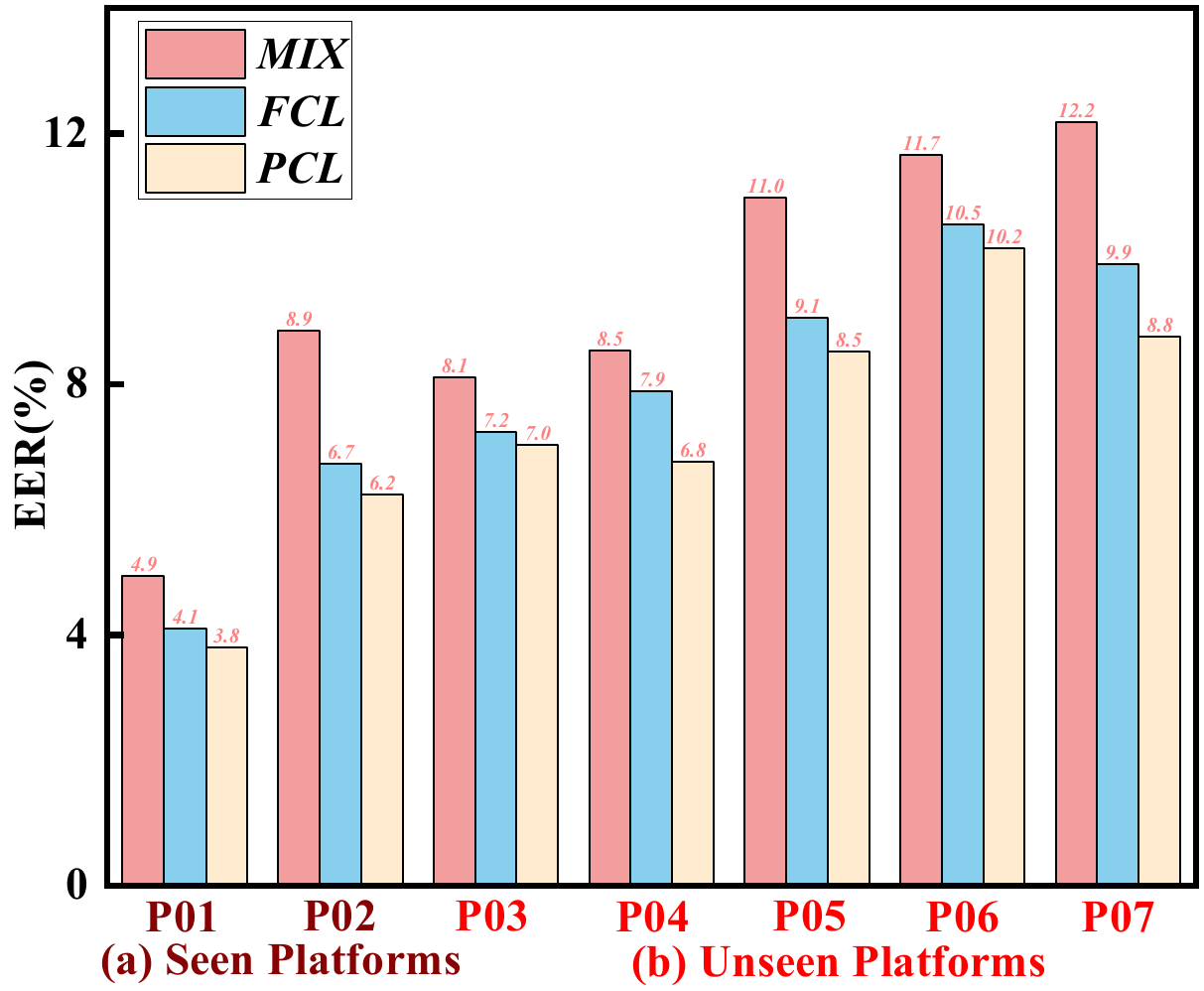}
	\caption{Performance comparison of mix training (MIX), frame-level consistency learning (FCL), and phoneme-level consistency learning (PCL) across seen and unseen communication platforms.}
	\label{fig:platform}
\end{figure}

Fig.~\ref{fig:platform} presents a comparison of cross-platform generalization performance, where (a) and (b) correspond to platforms seen and unseen during training, respectively. Three training strategies are compared: a baseline using mixed offline and online data (MIX), frame-level consistency learning (FCL), and phoneme-level consistency learning (PCL). This experiment is designed to evaluate the impact of different consistency modeling granularities on model generalization under cross-platform conditions.

As shown in Fig.~\ref{fig:platform}, both FCL and PCL consistently outperform the MIX baseline on seen platforms, demonstrating the effectiveness of consistency learning in mitigating the distribution discrepancy between offline and online data. Notably, PCL achieves lower EERs than FCL across all seen platforms, substantiating the prior analysis that phoneme-level modeling provides more stable representation alignment than frame-level constraints.

Under unseen platform conditions, the performance gap among the methods becomes even more pronounced. The MIX strategy suffers from severe performance degradation, particularly on platforms P05--P07, where EERs increase significantly. While FCL alleviates this degradation to some extent, it still exhibits notable platform-wise variability. In contrast, PCL maintains the lowest and most stable EERs across all unseen platforms. Its advantages are particularly evident on platforms with more severe distortions (e.g., P07). These results suggest that by mining robust features within linguistic structures, phoneme-level consistency learning enables the model to learn platform-invariant representations, thereby effectively enhancing generalization in real-world, complex communication environments.

\subsection{Robustness under Noise Scenarios}

Table~\ref{tab:eer_avg_scenarios} evaluates the robustness of different models under unseen noise conditions, where S01 corresponds to the clean condition and S02–S07 represent various types of unseen noise and interference scenarios. The results indicate that performance differences among training strategies become more pronounced under these unseen conditions, reflecting varying levels of generalization capability to complex real-world environments. It is worth noting that, in RTC scenarios, speech degradation is not only caused by environmental noise itself but is also significantly influenced by built-in speech enhancement modules of communication platforms, such as noise suppression, echo cancellation, and automatic gain control. The coupling between noise and enhancement processing introduces more complex and nonlinear distortion patterns.

By comparing single-source training strategies (Off / On) with mixed-data training (Mix), we observe that simple data mixing can partially alleviate distribution shifts under unseen conditions, leading to lower EERs in some noise scenarios. However, these improvements are not consistent across different unseen noise conditions, suggesting that relying solely on data-level mixing is insufficient to effectively model the complex distortion responses introduced by speech enhancement in real-time communication systems.

\begin{table}[t]
		\begin{threeparttable}
	\centering
	\caption{EER (\%) results under different evaluation conditions, averaged over offline and online settings. Details of S0X conditions are provided in Table~\ref{tab:gen_plat_ID}.}
	\label{tab:eer_avg_scenarios}
	\setlength{\tabcolsep}{1pt}
	\renewcommand{\arraystretch}{1.3}
	\begin{tabular}{c|c|cccccc}
		\specialrule{1.2pt}{0pt}{0pt}
		\multirow{2}{*}{\textbf{Train}}
		& \textbf{Seen} 
		& \multicolumn{6}{c}{\textbf{Unseen}} \\
		\cline{2-8}
		& S01$^{*}$
		& S02 
		& S03 
		& S04 
		& S05 
		& S06 
		& S07 \\
		\hline
		Off 
		& 7.68 & 17.24 & 16.05 & 16.56 & 18.65 & 14.28 & 15.28 \\
		On 
		& 6.66 & 12.33 & 12.60 & 17.34 & 14.30 & 11.27 & 11.92 \\
		Mix 
		& 5.63 & 12.80 & 12.72 & 16.92 & 13.61 & 12.11 & 10.80 \\
		\cellcolor{lightgray}\textbf{PCL}
		& \cellcolor{lightgray}\textbf{3.88}
		& \cellcolor{lightgray}\textbf{10.95}
		& \cellcolor{lightgray}\textbf{9.30}
		& \cellcolor{lightgray}\textbf{13.40}
		& \cellcolor{lightgray}\textbf{13.09}
		& \cellcolor{lightgray}\textbf{9.57}
		& \cellcolor{lightgray}\textbf{9.53} \\
		\specialrule{1.2pt}{0pt}{0pt}
	\end{tabular}
		\begin{tablenotes}
		\footnotesize \item $^{*}$S01 denotes clean-only.
		
	\end{tablenotes}
	\end{threeparttable}
\end{table}

In contrast, the PCL method demonstrates more stable performance across unseen noise scenarios. By enforcing consistency between offline and online representations at the phoneme level, PCL effectively suppresses structural shifts induced by speech enhancement modules, achieving a better balance between environmental noise variations and platform-specific processing distortions, and thereby significantly improving robustness under unseen real-time communication noise conditions.

\subsection{Ablation Studies}

In this subsection, we conduct ablation studies to validate the effectiveness of the proposed method.

\paragraph{Feature and Training Strategy Analysis.}
As shown in Table~\ref{tab:feat_strategy_ablation}, we compare different combinations of feature granularity and consistency strategies. The results indicate that frame-level features provide a stronger performance baseline than phoneme-level features. From the perspective of training strategies,  PCL significantly outperforms frame-level constraints. These findings suggest that preserving fine-grained low-level features while leveraging PCL to capture structured semantic information is more effective for RTC scenarios.

\begin{figure}[!t]  
	\centering
	\includegraphics[scale=0.43]{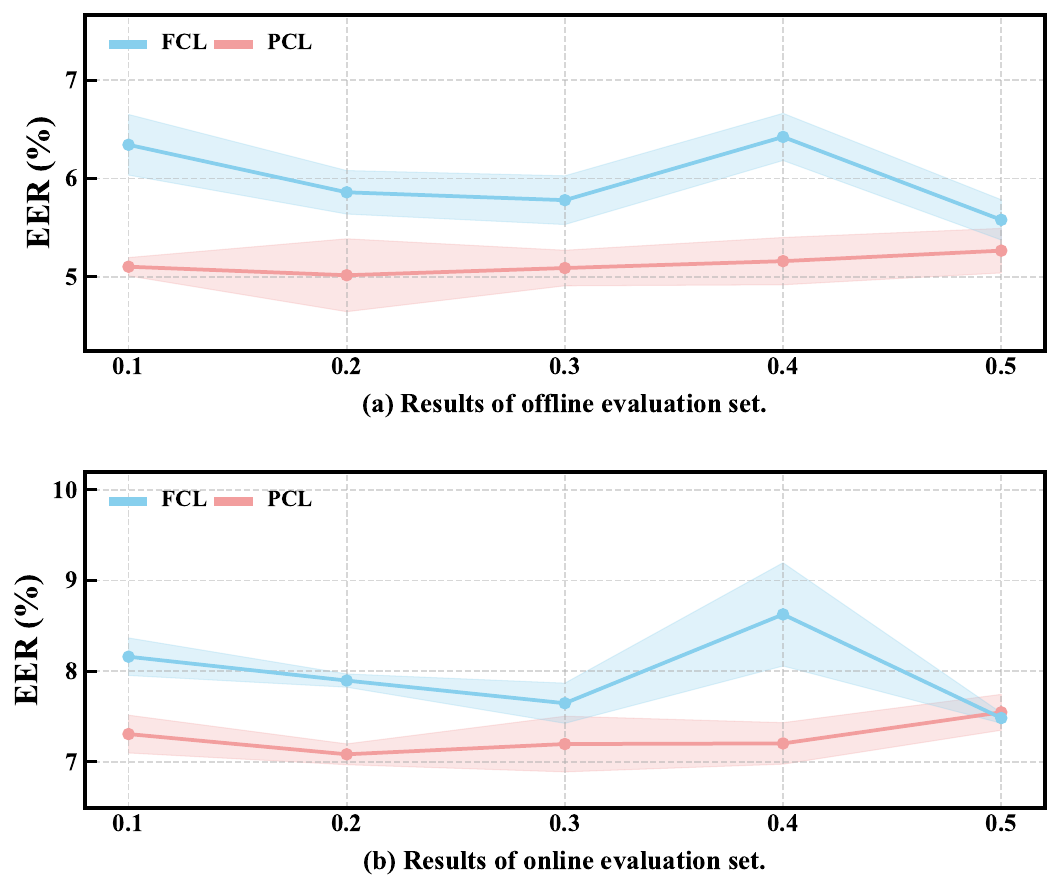}
	\caption{Comparison of FCL and PCL consistency learning on offline and online evaluation sets.
		Colored curves denote the mean EER across multiple runs under different values of the regularization weight $\lambda$, while the shaded regions indicate the corresponding minimum–maximum range.}
	\label{fig:pcl}
\end{figure}

\paragraph{Weight and Stability Analysis.}
Based on the above observations, Fig.~\ref{fig:pcl} further investigates the stability of different consistency strategies under varying weighting factors $\lambda$. Overall, PCL consistently outperforms FCL across all evaluation sets. Notably, the performance of PCL exhibits smaller fluctuations with respect to changes in $\lambda$, and the variance of the experimental results is substantially reduced. This demonstrates that phoneme-level constraints provide a more stable and robust regularization signal.

\begin{table}[t]
	\centering
	\caption{Ablation study of feature granularity and training strategies, where EER results are averaged over offline and online evaluation settings.}
	\label{tab:feat_strategy_ablation}
	\setlength{\tabcolsep}{4pt}
	
	\newcommand{\cmark}{\ding{51}} 
	\newcommand{\xmark}{\ding{55}} 
	
	\begin{tabular}{ccccc}
		\specialrule{1.2pt}{0pt}{0pt}
		\multicolumn{2}{c}{\textbf{Feature}} &
		\multicolumn{2}{c}{\textbf{Strategy}} &
		\multirow{2}{*}{EER (\%)} \\
		\cmidrule(lr){1-2}\cmidrule(lr){3-4}
		Frame & Phoneme &
		FCL & PCL &  \\
		\midrule
		
		\xmark & \cmark & \cmark & \xmark & 8.34 \\
		
		\textbf{\xmark} & \textbf{\cmark} & \textbf{\xmark} & \textbf{\cmark} & 7.52 \\
		\cmark & \xmark & \cmark & \xmark & 6.55 \\
		\rowcolor{gray!18}
		\cmark & \xmark & \xmark & \cmark & \textbf{5.81} \\
		
		\specialrule{1.2pt}{0pt}{0pt}
	\end{tabular}
\end{table}

\section{Conclusion}
This paper presents RTCFake, a pioneering speech deepfake detection dataset specifically constructed to investigate black-box transmission conditions within RTC scenarios. Through in-depth analysis, we reveal that while the coupled nonlinear processing in RTC systems severely distorts fine-grained acoustic details, phoneme-level representations maintain superior structural stability. Motivated by this, we propose a phoneme-guided consistency learning method that enforces alignment between offline and online representations at the semantic structural level, effectively mitigating the loss of discriminative cues during transmission. Extensive experiments demonstrate that our approach significantly enhances detection robustness and generalization across both seen and unseen communication platforms, as well as under diverse noise conditions. This work provides a solid data foundation and a robust methodological framework for deploying speech deepfake detection systems in realistic and complex communication environments.

\section{Acknowledgement}
This work is supported by the Natural Science Foundation of China (NSFC) under the grant NO.62572358, 62571002

\section*{Limitations}
Despite constructing the RTCFake dataset and proposing the phoneme-guided consistency learning method, this work has several limitations. First, real-world communication involves a vast array of confounding factors beyond transmission, such as the heterogeneity of recording/playback hardware and diverse user behaviors. The complex interplay between these terminal-side variables and the platforms' internal processing pipelines may introduce additional signal perturbations not fully captured in this study. Moreover, while our phoneme-level constraints demonstrate superior stability, a performance gap still exists when encountering extreme unseen noise or highly aggressive nonlinear distortions in certain communication platforms. Future research will focus on developing platform-agnostic and more adaptive representation modeling to further bridge the gap between laboratory evaluation and large-scale real-world deployment.


\section*{Ethics Statement}

The real speech samples used in the RTCFake dataset are sourced from publicly available speech corpora that are commonly employed in speech processing research. Based on these samples, deepfake speech is generated using text-to-speech (TTS) and voice conversion (VC) techniques, and subsequently transmitted through mainstream real-time communication platforms to construct online speech samples. The dataset does not include speech from identifiable real individuals, nor does it contain any harmful, sensitive, or privacy-related content.

\section*{Generative AI Use Disclosure}
Generative AI tools were used exclusively for language refinement and grammatical correction during the preparation of this manuscript.

\bibliography{custom}

\begin{thebibliography}{41}
\providecommand{\natexlab}[1]{#1}

\bibitem[{vol(2025)}]{volcengine}
 2025.
\newblock \url{https://www.volcengine.com/docs/6561/1257584?lang=en}.

\bibitem[{Babu et~al.(2022)Babu, Wang, Tjandra, Lakhotia, Xu, Goyal, Singh, von
  Platen, Saraf, Pino et~al.}]{babu2022xls}
Arun Babu, Changhan Wang, Andros Tjandra, Kushal Lakhotia, Qiantong Xu, Naman
  Goyal, Kritika Singh, Patrick von Platen, Yatharth Saraf, Juan Pino, and 1
  others. 2022.
\newblock Xls-r: Self-supervised cross-lingual speech representation learning
  at scale.
\newblock In \emph{Proc. Interspeech 2022}, pages 2278--2282.

\bibitem[{{Channel News Asia}(2025)}]{cna_deepfake_scam_2025}
{Channel News Asia}. 2025.
\newblock \href
  {https://www.channelnewsasia.com/singapore/deepfake-scam-impersonate-ceo-company-finance-director-5048706}
  {Company finance director nearly loses over {US}\$499{,}000 to scammers using
  deepfake to impersonate ceo}.
\newblock Accessed: 2025-11-6.

\bibitem[{Chen et~al.(2025)Chen, Niu, Ma, Deng, Wang, JianZhao, Yu, and
  Chen}]{chen2025f5}
Yushen Chen, Zhikang Niu, Ziyang Ma, Keqi Deng, Chunhui Wang, JianZhao
  JianZhao, Kai Yu, and Xie Chen. 2025.
\newblock F5-tts: A fairytaler that fakes fluent and faithful speech with flow
  matching.
\newblock In \emph{Proceedings of the 63rd Annual Meeting of the Association
  for Computational Linguistics (Volume 1: Long Papers)}, pages 6255--6271.

\bibitem[{Du et~al.(2024{\natexlab{a}})Du, Lin, Chiu, Chen, Wu, Ren, Tsao, Lee,
  and Jang}]{du2024dfadd}
Jiawei Du, I-Ming Lin, I-Hsiang Chiu, Xuanjun Chen, Haibin Wu, Wenze Ren,
  Yu~Tsao, Hung-yi Lee, and Jyh-Shing~Roger Jang. 2024{\natexlab{a}}.
\newblock Dfadd: The diffusion and flow-matching based audio deepfake dataset.
\newblock In \emph{2024 IEEE Spoken Language Technology Workshop (SLT)}, pages
  921--928. IEEE.

\bibitem[{Du et~al.(2024{\natexlab{b}})Du, Chen, Zhang, Hu, Lu, Yang, Hu,
  Zheng, Gu, Ma et~al.}]{du2024cosyvoice}
Zhihao Du, Qian Chen, Shiliang Zhang, Kai Hu, Heng Lu, Yexin Yang, Hangrui Hu,
  Siqi Zheng, Yue Gu, Ziyang Ma, and 1 others. 2024{\natexlab{b}}.
\newblock Cosyvoice: A scalable multilingual zero-shot text-to-speech
  synthesizer based on supervised semantic tokens.
\newblock \emph{arXiv preprint arXiv:2407.05407}.

\bibitem[{Fan et~al.(2024{\natexlab{a}})Fan, Ding, Tao, Fu, Yi, Wen, and
  Lv}]{fan2024dual}
Cunhang Fan, Mingming Ding, Jianhua Tao, Ruibo Fu, Jiangyan Yi, Zhengqi Wen,
  and Zhao Lv. 2024{\natexlab{a}}.
\newblock Dual-branch knowledge distillation for noise-robust synthetic speech
  detection.
\newblock \emph{IEEE/ACM Transactions on Audio, Speech, and Language
  Processing}, 32:2453--2466.

\bibitem[{Fan et~al.(2024{\natexlab{b}})Fan, Xue, Tao, Yi, Wang, Zheng, and
  Lv}]{fan2024spatial}
Cunhang Fan, Jun Xue, Jianhua Tao, Jiangyan Yi, Chenglong Wang, Chengshi Zheng,
  and Zhao Lv. 2024{\natexlab{b}}.
\newblock Spatial reconstructed local attention res2net with f0 subband for
  fake speech detection.
\newblock \emph{Neural Networks}, 175:106320.

\bibitem[{Huang et~al.(2025)Huang, Gu, Wang, Zhu, and Qian}]{wen2025speechfake}
Wen Huang, Yanmei Gu, Zhiming Wang, Huijia Zhu, and Yanmin Qian. 2025.
\newblock Speechfake: A large-scale multilingual speech deepfake dataset
  incorporating cutting-edge generation methods.
\newblock In \emph{Proceedings of the 63rd Annual Meeting of the Association
  for Computational Linguistics (Volume 1: Long Papers)}, page 9985–9998.
  Association for Computational Linguistics.

\bibitem[{Jung et~al.(2022)Jung, Heo, Tak, Shim, Chung, Lee, Yu, and
  Evans}]{jung2022aasist}
Jee-weon Jung, Hee-Soo Heo, Hemlata Tak, Hye-jin Shim, Joon~Son Chung, Bong-Jin
  Lee, Ha-Jin Yu, and Nicholas Evans. 2022.
\newblock Aasist: Audio anti-spoofing using integrated spectro-temporal graph
  attention networks.
\newblock In \emph{ICASSP 2022-2022 IEEE international conference on acoustics,
  speech and signal processing (ICASSP)}, pages 6367--6371. IEEE.

\bibitem[{Jung et~al.(2025)Jung, Wu, Wang, Kim, Maiti, Matsunaga, Shim, Tian,
  Evans, Chung et~al.}]{jung2025spoofceleb}
Jee-weon Jung, Yihan Wu, Xin Wang, Ji-Hoon Kim, Soumi Maiti, Yuta Matsunaga,
  Hye-jin Shim, Jinchuan Tian, Nicholas Evans, Joon~Son Chung, and 1 others.
  2025.
\newblock Spoofceleb: Speech deepfake detection and sasv in the wild.
\newblock \emph{IEEE Open Journal of Signal Processing}.

\bibitem[{Kang et~al.(2024)Kang, Yang, Yao, Kuang, Yang, Guo, Lin, and
  Povey}]{kang2024libriheavy}
Wei Kang, Xiaoyu Yang, Zengwei Yao, Fangjun Kuang, Yifan Yang, Liyong Guo, Long
  Lin, and Daniel Povey. 2024.
\newblock Libriheavy: A 50,000 hours asr corpus with punctuation casing and
  context.
\newblock In \emph{ICASSP 2024-2024 IEEE International Conference on Acoustics,
  Speech and Signal Processing (ICASSP)}, pages 10991--10995. IEEE.

\bibitem[{Li et~al.(2024{\natexlab{a}})Li, Li, Zheng, Yan, Ji, and
  Xu}]{li2024safeear}
Xinfeng Li, Kai Li, Yifan Zheng, Chen Yan, Xiaoyu Ji, and Wenyuan Xu.
  2024{\natexlab{a}}.
\newblock Safeear: Content privacy-preserving audio deepfake detection.
\newblock In \emph{Proceedings of the 2024 on ACM SIGSAC Conference on Computer
  and Communications Security}, pages 3585--3599.

\bibitem[{Li et~al.(2024{\natexlab{b}})Li, Zhang, Ren, Qiao, Ma, Wei, and
  Yang}]{li2024cross}
Yuang Li, Min Zhang, Mengxin Ren, Xiaosong Qiao, Miaomiao Ma, Daimeng Wei, and
  Hao Yang. 2024{\natexlab{b}}.
\newblock Cross-domain audio deepfake detection: Dataset and analysis.
\newblock In \emph{Proceedings of the 2024 Conference on Empirical Methods in
  Natural Language Processing}, pages 4977--4983.

\bibitem[{Liao et~al.(2024)Liao, Wang, Li, Cheng, Zhang, Zhou, and
  Xing}]{liao2024fish}
Shijia Liao, Yuxuan Wang, Tianyu Li, Yifan Cheng, Ruoyi Zhang, Rongzhi Zhou,
  and Yijin Xing. 2024.
\newblock Fish-speech: Leveraging large language models for advanced
  multilingual text-to-speech synthesis.
\newblock \emph{arXiv preprint arXiv:2411.01156}.

\bibitem[{Liu(2024)}]{liu2024zero}
Songting Liu. 2024.
\newblock Zero-shot voice conversion with diffusion transformers.
\newblock \emph{arXiv preprint arXiv:2411.09943}.

\bibitem[{Liu et~al.(2023)Liu, Wang, Sahidullah, Patino, Delgado, Kinnunen,
  Todisco, Yamagishi, Evans, Nautsch et~al.}]{liu2023asvspoof}
Xuechen Liu, Xin Wang, Md~Sahidullah, Jose Patino, H{\'e}ctor Delgado, Tomi
  Kinnunen, Massimiliano Todisco, Junichi Yamagishi, Nicholas Evans, Andreas
  Nautsch, and 1 others. 2023.
\newblock Asvspoof 2021: Towards spoofed and deepfake speech detection in the
  wild.
\newblock \emph{IEEE/ACM Transactions on Audio, Speech, and Language
  Processing}, 31:2507--2522.

\bibitem[{M{\"u}ller et~al.(2024)M{\"u}ller, Kawa, Choong, Casanova, G{\"o}lge,
  M{\"u}ller, Syga, Sperl, and B{\"o}ttinger}]{muller2024mlaad}
Nicolas~M M{\"u}ller, Piotr Kawa, Wei~Herng Choong, Edresson Casanova, Eren
  G{\"o}lge, Thorsten M{\"u}ller, Piotr Syga, Philip Sperl, and Konstantin
  B{\"o}ttinger. 2024.
\newblock Mlaad: The multi-language audio anti-spoofing dataset.
\newblock In \emph{2024 International Joint Conference on Neural Networks
  (IJCNN)}, pages 1--7. IEEE.

\bibitem[{Piczak(2015)}]{piczak2015esc}
Karol~J Piczak. 2015.
\newblock Esc: Dataset for environmental sound classification.
\newblock In \emph{Proceedings of the 23rd ACM international conference on
  Multimedia}, pages 1015--1018.

\bibitem[{Radford et~al.(2023)Radford, Kim, Xu, Brockman, McLeavey, and
  Sutskever}]{radford2023robust}
Alec Radford, Jong~Wook Kim, Tao Xu, Greg Brockman, Christine McLeavey, and
  Ilya Sutskever. 2023.
\newblock Robust speech recognition via large-scale weak supervision.
\newblock In \emph{International conference on machine learning}, pages
  28492--28518. PMLR.

\bibitem[{{Resemble AI}(2025)}]{chatterboxtts2025}
{Resemble AI}. 2025.
\newblock {Chatterbox-TTS}.
\newblock \url{https://github.com/resemble-ai/chatterbox}.
\newblock GitHub repository.

\bibitem[{Tak et~al.(2022{\natexlab{a}})Tak, Kamble, Patino, Todisco, and
  Evans}]{tak2022rawboost}
Hemlata Tak, Madhu Kamble, Jose Patino, Massimiliano Todisco, and Nicholas
  Evans. 2022{\natexlab{a}}.
\newblock Rawboost: A raw data boosting and augmentation method applied to
  automatic speaker verification anti-spoofing.
\newblock In \emph{ICASSP 2022-2022 IEEE International Conference on Acoustics,
  Speech and Signal Processing (ICASSP)}, pages 6382--6386. IEEE.

\bibitem[{Tak et~al.(2022{\natexlab{b}})Tak, Todisco, Wang, Jung, Yamagishi,
  and Evans}]{tak2022automatic}
Hemlata Tak, Massimiliano Todisco, Xin Wang, Jee-weon Jung, Junichi Yamagishi,
  and Nicholas Evans. 2022{\natexlab{b}}.
\newblock Automatic speaker verification spoofing and deepfake detection using
  wav2vec 2.0 and data augmentation.
\newblock In \emph{The Speaker and Language Recognition Workshop (Odyssey
  2022)}. ISCA.

\bibitem[{Todisco et~al.(2019)Todisco, Wang, Vestman, Sahidullah, Delgado,
  Nautsch, Yamagishi, Evans, Kinnunen, and Lee}]{todisco2019asvspoof}
Massimiliano Todisco, Xin Wang, Ville Vestman, Md~Sahidullah, Hector Delgado,
  Andreas Nautsch, Junichi Yamagishi, Nicholas Evans, Tomi Kinnunen, and
  Kong~Aik Lee. 2019.
\newblock Asvspoof 2019: Future horizons in spoofed and fake audio detection.
\newblock In \emph{Interspeech 2019}, pages 1008--1012. International Speech
  Communication Association.

\bibitem[{Wang et~al.(2024)Wang, Delgado, Tak, Jung, Shim, Todisco, Kukanov,
  Liu, Sahidullah, Kinnunen et~al.}]{wang2024asvspoof}
Xin Wang, H{\'e}ctor Delgado, Hemlata Tak, Jee-Weon Jung, Hye-Jin Shim,
  Massimiliano Todisco, Ivan Kukanov, Xuechen Liu, Md~Sahidullah, Tomi
  Kinnunen, and 1 others. 2024.
\newblock Asvspoof 5: crowdsourced speech data, deepfakes, and adversarial
  attacks at scale.
\newblock In \emph{The Automatic Speaker Verification Spoofing Countermeasures
  Workshop (ASVspoof 2024)}, pages 1--8. ISCA.

\bibitem[{Wang et~al.(2025)Wang, Jiang, Ma, Zhang, Liu, Li, Liang, Zheng, Wang,
  Feng et~al.}]{wang2025spark}
Xinsheng Wang, Mingqi Jiang, Ziyang Ma, Ziyu Zhang, Songxiang Liu, Linqin Li,
  Zheng Liang, Qixi Zheng, Rui Wang, Xiaoqin Feng, and 1 others. 2025.
\newblock Spark-tts: An efficient llm-based text-to-speech model with
  single-stream decoupled speech tokens.
\newblock \emph{arXiv preprint arXiv:2503.01710}.

\bibitem[{Wu et~al.(2024)Wu, Tseng, and Lee}]{wu2024codecfake}
Haibin Wu, Yuan Tseng, and Hung-yi Lee. 2024.
\newblock Codecfake: Enhancing anti-spoofing models against deepfake audios
  from codec-based speech synthesis systems.
\newblock In \emph{Proc. Interspeech 2024}, pages 1770--1774.

\bibitem[{Xie et~al.(2025{\natexlab{a}})Xie, Fu, Wang, Wang, Li, Wen, Cheng,
  and Ye}]{xie2025fake}
Yuankun Xie, Ruibo Fu, Xiaopeng Wang, Zhiyong Wang, Ya~Li, Zhengqi Wen, Haonnan
  Cheng, and Long Ye. 2025{\natexlab{a}}.
\newblock Fake speech wild: Detecting deepfake speech on social media platform.
\newblock \emph{arXiv preprint arXiv:2508.10559}.

\bibitem[{Xie et~al.(2025{\natexlab{b}})Xie, Lu, Fu, Wen, Wang, Tao, Qi, Wang,
  Liu, Cheng et~al.}]{xie2025codecfake}
Yuankun Xie, Yi~Lu, Ruibo Fu, Zhengqi Wen, Zhiyong Wang, Jianhua Tao, Xin Qi,
  Xiaopeng Wang, Yukun Liu, Haonan Cheng, and 1 others. 2025{\natexlab{b}}.
\newblock The codecfake dataset and countermeasures for the universally
  detection of deepfake audio.
\newblock \emph{IEEE Transactions on Audio, Speech and Language Processing}.

\bibitem[{Xu et~al.(2022)Xu, Baevski, and Auli}]{xu2022simple}
Qiantong Xu, Alexei Baevski, and Michael Auli. 2022.
\newblock Simple and effective zero-shot cross-lingual phoneme recognition.
\newblock In \emph{Proc. Interspeech 2022}, pages 2113--2117.

\bibitem[{Xue et~al.(2022)Xue, Fan, Lv, Tao, Yi, Zheng, Wen, Yuan, and
  Shao}]{xue2022audio}
Jun Xue, Cunhang Fan, Zhao Lv, Jianhua Tao, Jiangyan Yi, Chengshi Zheng,
  Zhengqi Wen, Minmin Yuan, and Shegang Shao. 2022.
\newblock Audio deepfake detection based on a combination of f0 information and
  real plus imaginary spectrogram features.
\newblock In \emph{Proceedings of the 1st international workshop on deepfake
  detection for audio multimedia}, pages 19--26.

\bibitem[{Xue et~al.(2023)Xue, Fan, Yi, Wang, Wen, Zhang, and
  Lv}]{xue2023learning}
Jun Xue, Cunhang Fan, Jiangyan Yi, Chenglong Wang, Zhengqi Wen, Dan Zhang, and
  Zhao Lv. 2023.
\newblock Learning from yourself: A self-distillation method for fake speech
  detection.
\newblock In \emph{ICASSP 2023-2023 IEEE International Conference on Acoustics,
  Speech and Signal Processing (ICASSP)}, pages 1--5. IEEE.

\bibitem[{Xue et~al.(2024)Xue, Fan, Yi, Zhou, and Lv}]{xue2024dynamic}
Jun Xue, Cunhang Fan, Jiangyan Yi, Jian Zhou, and Zhao Lv. 2024.
\newblock Dynamic ensemble teacher-student distillation framework for
  light-weight fake audio detection.
\newblock \emph{IEEE Signal Processing Letters}, 31:2305--2309.

\bibitem[{Yamagishi et~al.(2021)Yamagishi, Wang, Todisco, Sahidullah, Patino,
  Nautsch, Liu, Lee, Kinnunen, Evans et~al.}]{yamagishi2021asvspoof}
Junichi Yamagishi, Xin Wang, Massimiliano Todisco, Md~Sahidullah, Jose Patino,
  Andreas Nautsch, Xuechen Liu, Kong~Aik Lee, Tomi Kinnunen, Nicholas Evans,
  and 1 others. 2021.
\newblock Asvspoof 2021: accelerating progress in spoofed and deepfake speech
  detection.
\newblock In \emph{Proc. ASVSPOOF 2021}, pages 47--54.

\bibitem[{Ye et~al.(2025)Ye, Zhu, Chan, Wang, Tan, Lei, Peng, Liu, Jin, Dai
  et~al.}]{ye2025llasa}
Zhen Ye, Xinfa Zhu, Chi-Min Chan, Xinsheng Wang, Xu~Tan, Jiahe Lei, Yi~Peng,
  Haohe Liu, Yizhu Jin, Zheqi Dai, and 1 others. 2025.
\newblock Llasa: Scaling train-time and inference-time compute for llama-based
  speech synthesis.
\newblock \emph{arXiv preprint arXiv:2502.04128}.

\bibitem[{Yi et~al.(2022)Yi, Fu, Tao, Nie, Ma, Wang, Wang, Tian, Bai, Fan
  et~al.}]{yi2022add}
Jiangyan Yi, Ruibo Fu, Jianhua Tao, Shuai Nie, Haoxin Ma, Chenglong Wang, Tao
  Wang, Zhengkun Tian, Ye~Bai, Cunhang Fan, and 1 others. 2022.
\newblock Add 2022: the first audio deep synthesis detection challenge.
\newblock In \emph{ICASSP 2022-2022 IEEE International Conference on Acoustics,
  Speech and Signal Processing (ICASSP)}, pages 9216--9220. IEEE.

\bibitem[{Yi et~al.(2023)Yi, Tao, Fu, Yan, Wang, Wang, Zhang, Zhang, Zhao, Ren
  et~al.}]{yi2023add}
Jiangyan Yi, Jianhua Tao, Ruibo Fu, Xinrui Yan, Chenglong Wang, Tao Wang,
  Chu~Yuan Zhang, Xiaohui Zhang, Yan Zhao, Yong Ren, and 1 others. 2023.
\newblock Add 2023: the second audio deepfake detection challenge.
\newblock \emph{arXiv preprint arXiv:2305.13774}.

\bibitem[{Zhang et~al.(2024)Zhang, Wen, and Hu}]{zhang2024audio}
Qishan Zhang, Shuangbing Wen, and Tao Hu. 2024.
\newblock Audio deepfake detection with self-supervised xls-r and sls
  classifier.
\newblock In \emph{Proceedings of the 32nd ACM International Conference on
  Multimedia}, pages 6765--6773.

\bibitem[{Zhao et~al.(2025)Zhao, Jia, Wang, Zhou, Wang, and
  Qin}]{zhao2025chinese}
Jinghua Zhao, Yuhang Jia, Shiyao Wang, Jiaming Zhou, Hui Wang, and Yong Qin.
  2025.
\newblock Chinese-lips: A chinese audio-visual speech recognition dataset with
  lip-reading and presentation slides.
\newblock \emph{arXiv preprint arXiv:2504.15066}.

\bibitem[{Zhou et~al.(2025{\natexlab{a}})Zhou, Zhou, He, Zhou, Wang, Deng, and
  Shu}]{zhou2025indextts2}
Siyi Zhou, Yiquan Zhou, Yi~He, Xun Zhou, Jinchao Wang, Wei Deng, and Jingchen
  Shu. 2025{\natexlab{a}}.
\newblock Indextts2: A breakthrough in emotionally expressive and
  duration-controlled auto-regressive zero-shot text-to-speech.
\newblock \emph{arXiv preprint arXiv:2506.21619}.

\bibitem[{Zhou et~al.(2025{\natexlab{b}})Zhou, Zeng, Liu, Li, Yu, Wang, Ye,
  Sun, Gui, Li et~al.}]{zhou2025voxcpm}
Yixuan Zhou, Guoyang Zeng, Xin Liu, Xiang Li, Renjie Yu, Ziyang Wang, Runchuan
  Ye, Weiyue Sun, Jiancheng Gui, Kehan Li, and 1 others. 2025{\natexlab{b}}.
\newblock Voxcpm: Tokenizer-free tts for context-aware speech generation and
  true-to-life voice cloning.
\newblock \emph{arXiv preprint arXiv:2509.24650}.

\end{thebibliography}

\appendix

\section{Dataset Details}
\label{sec:appendix}

\begin{table*}[!h]
	\centering
	\caption{Configuration of speech generation methods, transmission platforms, and noise conditions.}
	\label{tab:gen_plat_ID}
	\begin{tabular}{|c|c|c|c|c|c|c|}
		\specialrule{1.2pt}{0pt}{0pt}
		\textbf{ID} & \textbf{Type} & \textbf{Generation} & \textbf{ID} & \textbf{Platform} & \textbf{ID} & \textbf{Noise} \\
		\hline
		G01 & TTS & F5-TTS \cite{chen2025f5} & P01 & Zoom & S01 & Clean \\
		\hline
		G02 & TTS & OpenAudio-S1 \cite{liao2024fish} & P02 & QQ & S02 & Office \\
		\hline
		G03 & TTS & VOXCPM \cite{zhou2025voxcpm} & P03 & Wechat & S03 & Coffee \\
		\hline
		G04 & TTS & LLaSA \cite{ye2025llasa} & P04 & Dingtalk & S04 & Echo \\
		\hline
		G05 & TTS & IndexTTS2 \cite{zhou2025indextts2} & P05 & Lark & S05 & Rain \\
		\hline
		G06 & TTS & Doubao \cite{volcengine} & P06 & Voov & S06 & Footsteps \\
		\hline
		G07 & TTS & SparkTTS \cite{wang2025spark} & P07 & Telegram & S07 &Keyboard \\
		\hline
		G08 & VC & CosyVoice \cite{du2024cosyvoice} & -- & -- & -- & -- \\
		\hline
		G09 & VC & SeedVC \cite{liu2024zero} & -- & -- & -- & -- \\
		\hline
		G10 & VC & ChatterboxVC \cite{chatterboxtts2025} & -- & -- & -- & -- \\
		\specialrule{1.2pt}{0pt}{0pt}
	\end{tabular}
\end{table*}

\subsection{Dataset Partition}
The construction and partition of the dataset in this study involve multiple speech generation models, RTC platforms, and noise conditions. To facilitate a unified description of the experimental setup, simplify subsequent statistical analysis and result reporting, and avoid repetitive and lengthy naming in tables and the main text, we assign unified identifiers to these components and perform systematic dataset partitioning and statistics based on this encoding scheme.

Table \ref{tab:gen_plat_ID} summarizes the overall configuration of the speech generation methods, transmission platforms, and noise conditions used in this work. Specifically, different speech generation models, including TTS and VC approaches, are denoted as G01–G10; mainstream real-time communication platforms are denoted as P01–P07; and different noise or interference scenarios are denoted as S01–S07. This identifier scheme is consistently used throughout the experimental setup, dataset partition description, and result analysis to improve conciseness and readability.

Based on this configuration, Table \ref{tab:offline_dataset_overview} presents the partitioning setup and speaker gender statistics of the RTCFake dataset across the training (Train), validation (Dev), and evaluation (Eval) subsets. For each subset, the included ranges of speech generation models (denoted by Gxx identifiers) and the corresponding platform ranges (Pxx) are explicitly specified, together with the numbers of female and male speakers. This table aims to provide an intuitive comparison of the composition and scale differences among subsets of the offline dataset from the perspectives of speaker distribution and generation method coverage.

Furthermore, Table \ref{tab:real_fake_stats} provides a comprehensive statistical summary of the sample counts for both bonafide and fake speech under offline and online settings. We report the number of samples for the training, development, and evaluation phases, along with their respective totals. These statistics reveal the scale, class imbalance, and distribution characteristics of the data under different environmental constraints, providing essential context for interpreting the performance benchmarks in the following sections.

Overall, through the joint presentation of these Tables, we systematically characterize the construction elements, partitioning strategies, and scale distribution of our dataset, thereby providing a solid foundation for rigorous result analysis.

\begin{table}[t]
	\centering
	\caption{Subset configuration and speaker gender statistics of the RTCFake dataset.}
	\label{tab:offline_dataset_overview}
	\setlength{\tabcolsep}{1pt}
	
	\begin{tabular}{c c c c c}
		\specialrule{1.2pt}{0pt}{0pt}
		\textbf{Subset} 
		& \textbf{\#Gen} 
		& \textbf{\#Platform}
		& \textbf{\#Female} 
		& \textbf{\#Male} \\
		\midrule
		
		\multirow{2}{*}{Train} 
		& G01--G04 
		& \multirow{2}{*}{P01--P02}
		& \multirow{2}{*}{55} 
		& \multirow{2}{*}{50} \\
		& G08--G09 
		&  &  &  \\
		\midrule
		
		\multirow{2}{*}{Dev} 
		& G01--G04 
		& \multirow{2}{*}{P01--P03}
		& \multirow{2}{*}{10} 
		& \multirow{2}{*}{11} \\
		& G08--G09 
		&  &  &  \\
		\midrule
		
		\multirow{2}{*}{Eval} 
		& G01--G07 
		& \multirow{2}{*}{P01--P07}
		& \multirow{2}{*}{95} 
		& \multirow{2}{*}{86} \\
		& G08--G10 
		&  &  &  \\
		
		\specialrule{1.2pt}{0pt}{0pt}
	\end{tabular}
\end{table}

\subsection{Dataset Metadata}

RTCFake provides comprehensive metadata for each speech sample to support flexible dataset partitioning, experimental analysis, and reproducibility. The metadata associated with each sample includes the following components:

\begin{itemize}
	\item \textbf{Basic Labels:} Indicating whether the speech sample is real or fake.
	
	\item \textbf{Generation Method:} Specifying the speech generation approach used to create the sample, including TTS and VC models. Each generation method is represented using a unified identifier (e.g., G01--G10).
	
	\item \textbf{RTC Platform:} Samples in the online subset are annotated with the RTC platform through which they are transmitted, such as Zoom, QQ, or WeChat. Different platforms are encoded using platform identifiers (e.g., P01–P07).
	
	\item \textbf{Speaker ID:} Providing a unique identity label for the speaker associated with the speech sample. For spoofed samples, the speaker ID corresponds to the target or source speaker used during speech generation.
	
	\item \textbf{Language ID:} Indicating the language of the audio sample.
	
	\item \textbf{Text Description:} Providing the corresponding textual transcription of the speech content, which serves as the input text for TTS generation or the reference content for speech analysis.
\end{itemize}

\begin{table}[t]
	\centering
	\caption{Statistics of real and fake samples under offline and online settings.}
	\label{tab:real_fake_stats}
	\setlength{\tabcolsep}{6pt}
	
	\begin{tabular}{c|cccc}
		\specialrule{1.2pt}{0pt}{0pt}
		\textbf{Set} 
		& \textbf{Train} 
		& \textbf{Dev} 
		& \textbf{Eval} 
		& \textbf{Total} \\
		\midrule
		
		\rowcolor{gray!12}
		\multicolumn{5}{l}{\footnotesize\textbf{\textit{Real Data}}} \\
		offline & 7463 & 1452 & 14948 & 23863 \\
		online  & 6890 & 1397 & 22737 & 31024 \\
		
		\midrule
		
		\rowcolor{gray!12}
		\multicolumn{5}{l}{\footnotesize\textbf{\textit{Fake Data}}} \\
		offline & 31197 & 6287 & 74806 & 112290 \\
		online  & 30235 & 6068 & 113596 & 149899 \\
		
		\specialrule{1.2pt}{0pt}{0pt}
	\end{tabular}
\end{table}

\subsection{Generation and Platforms}

\begin{table}[t]
	\centering
	\caption{Training configurations of the W2V+ AASIST model used in experiments.}
	\label{tab:train_config_w2v_aasist}
	\setlength{\tabcolsep}{2pt}
	\begin{tabular}{l c}
		\specialrule{1.2pt}{0pt}{0pt}
		\textbf{Configurations} & \textbf{W2V+ AASIST} \\
		\midrule
		Model Size & 3M \\
		Input Audio & 16K \\
		Data augmentation & Rawboost \\
		Optimizer & Adam \\
		Learning Rate & 1e-6 \\
		Weight Decay & 1e-4 \\
		Total Epochs & 100 \\
		Early Stopping & 10 epochs \\
		Loss Function & Cross Entropy \\
		Constraint Loss & Mean Squared Error  \\
		
		\specialrule{1.2pt}{0pt}{0pt}
	\end{tabular}
\end{table}

\begin{table*}[!tp]
	\centering
	\footnotesize
	\caption{Speech generation models and real-time communication (RTC) platforms used in this work.}
	\label{tab:gen_platform}
	\setlength{\tabcolsep}{2pt}
	\renewcommand{\arraystretch}{1.}
	\begin{tabular}{c l p{7.0cm}}
		\specialrule{1.2pt}{0pt}{0pt}
		
		\textbf{No.} & \textbf{Generation} & \textbf{Link} \\
		\midrule
		
		1  & F5-TTS~\cite{chen2025f5} 
		& \url{https://github.com/SWivid/F5-TTS} \\
		
		2  & OpenAudio-S1~\cite{liao2024fish} 
		& \url{https://github.com/fishaudio/fish-speech} \\
		
		3  & VOXCPM~\cite{zhou2025voxcpm} 
		& \url{https://github.com/OpenBMB/VoxCPM} \\
		
		4  & LLaSA~\cite{ye2025llasa} 
		& \url{https://huggingface.co/HKUSTAudio/Llasa-3B} \\
		
		5  & IndexTTS2~\cite{zhou2025indextts2} 
		& \url{https://github.com/index-tts/index-tts?tab=readme-ov-file} \\
		
		6  & Doubao~\cite{volcengine} 
		& \url{https://www.volcengine.com/docs/6561/1257584?lang=en} \\
		
		7  & SparkTTS~\cite{wang2025spark} 
		& \url{https://github.com/SparkAudio/Spark-TTS} \\
		
		8  & CosyVoice~\cite{du2024cosyvoice} 
		& \url{https://github.com/FunAudioLLM/CosyVoice} \\
		
		9  & SeedVC~\cite{liu2024zero} 
		& \url{https://github.com/Plachtaa/seed-vc} \\
		
		10 & ChatterboxVC~\cite{chatterboxtts2025} 
		& \url{https://github.com/resemble-ai/chatterbox} \\
		
		\midrule
		
		\textbf{No.} & \textbf{Platform} & \textbf{Link} \\
		\midrule
		
		1  & Zoom 
		& \url{https://www.zoom.com/} \\
		
		2  & QQ 
		& \url{https://im.qq.com/index/} \\
		
		3  & WeChat 
		& \url{https://www.wechat.com/} \\
		
		4  & DingTalk 
		& \url{https://www.dingtalk.com/en} \\
		
		5  & Lark 
		& \url{https://www.larksuite.com/en_sg/} \\
		
		6  & VooV 
		& \url{https://voovmeeting.com/} \\
		
		7  & Telegram 
		& \url{https://telegram.org/} \\
		
		\specialrule{1.2pt}{0pt}{0pt}
	\end{tabular}
\end{table*}

Table \ref{tab:gen_platform} summarizes the 10 speech generation tools and 7 mainstream RTC platforms used to construct the RTCFake dataset. These tools represent the state-of-the-art (SOTA) in both TTS and VC. Furthermore, the selected RTC platforms are widely utilized in social, professional, and educational scenarios, ensuring the practical relevance of the dataset.

%
\section{Experiment Details}
\subsection{Experiment Setting}
Table 8 summarizes the training configuration of the W2V+ AASIST \cite{tak2022automatic} model used in our experiments. We follow the standard training setup adopted in prior AASIST-based speech deepfake detection studies. The model takes 16 kHz audio as input and is optimized using the Adam optimizer with a learning rate of $1\times10^{-6}$ and a weight decay of $1\times10^{-4}$, with RawBoost \cite{tak2022rawboost} applied as the data augmentation strategy.
The model is trained for up to 100 epochs, with early stopping triggered if no improvement on the validation set is observed for 10 consecutive epochs. During training, cross-entropy loss is used for the classification objective, while mean squared error loss is employed for consistency learning.

\end{document}